\documentclass[prb,preprint,showpacs,preprintnumbers,amsmath,amssymb]{revtex4}

\usepackage{graphicx}
\usepackage{dcolumn}
\usepackage{bm}
\usepackage{amsmath}
\usepackage[mathscr]{eucal}

\begin{document}
\preprint{preprint  ver: 2.5}

\title[Cu4]{Revisiting the stable structure of the Cu$_{4}$ complex in silicon}

\author{Takayoshi Fujimura}
\author{Koun Shirai}
 \email{koun@sanken.osaka-u.ac.jp}
\affiliation{%
Nanoscience and Nanotechnology Center, The Institute of Scientific and Industrial Research (ISIR), Osaka University, 8-1 Mihogaoka, Ibaraki, Osaka 567-0047, Japan
}%

\date{\today}

\begin{abstract} 
The photoluminescence (PL) spectrum of Cu-containing silicon has a sharp zero-phonon (ZP) band at 1.014 eV. The luminescence center corresponding to this band is called Cu$_{\rm PL}$ and is known to have the local $C_{3v}$ symmetry. A recent measurement by ultrahigh-resolution PL spectroscopy revealed that the Cu$_{\rm PL}$ center is a Cu$_{4}$ complex. Later, it was shown, by first-principles calculations, that the structure was Cu$_{(s)}$Cu$_{3(i)}$, that is, a complex composed of three interstitial Cu$_{(i)}$ atoms around a substitutional Cu$_{(s)}$ atom. This complex (called $C$-type) has the desired symmetry.
However, in this study, we show that the lowest-energy structure is different. The tetrahedral structure Cu$_{4}$, called $T$-type, has the lowest energy, with the value being 0.26 eV lower than that of $C$-type. Between these two types, there is an energy barrier of 0.14 eV, which allows $C$-type to exist in a metastable state. Details of the electronic properties of the $T$-type complex are given, by comparing with $C$-type and other isovalent complexes such as Li$_{4}$. Whereas the Cu$_{4}$ tetrahedron is incorporated in silicon in a manner compatible with the tetrahedral network, it also has its own molecular orbitals that exhibit metallic characteristics, in contrast to other complexes.
The ZP of the PL spectrum is very likely ascribed to the backflow mode of the Cu$_{4}$ tetrahedron.
\end{abstract}

\maketitle

\section{Introduction}
\label{sec:intro}
In silicon technology, a copper defect is known as a fast diffusing species; a Cu atom migrates throughout the entire depth of a wafer in a short time, {\em e.g.}, a depth  of 0.8 cm for 1 h at temperature of 200$^{\circ}$C.\cite{Weber83,Istratov02,Myers00} Copper atoms mostly reside at tetrahedral interstitial sites, whereas they at sometimes jump to neighboring tetrahedral interstitial sites.\cite{Hall64,Meek75} The majority of Cu defects are of this type, which is known as interstitial Cu, denoted as Cu$_{(i)}$. Although interstitial Cu$_{(i)}$ donates an electron, it is electronically inactive in the sense that no gap state is observed.
However, many electronic signatures attributed to Cu defects have been found in a variety of measurements such as DLTS, \cite{Graff81,Lemke86,Brotherton87,Erzgraber95,Istratov97,Istratov98a,Nakamura09} photoluminescence (PL),\cite{Weber82} EPR,\cite{Hai97} photocurrent-induced DLTS,\cite{Brotherton87} Laplace-DLTS,\cite{Yarykin13a} IR spectroscopy,\cite{Teklemichael14} and others. \cite{Mesli92,Heiser97,Istratov98,Wahl00,Matsukawa19}
This indicates that there are many ways of incorporating a Cu atom in the host crystal. 
The simplest way of incorporation other than an interstitial type is the substitution of a host atom, denoted as Cu$_{(s)}$. However, even for the simplest type Cu$_{(s)}$, the identification of the electronic levels among many observed levels has not been accomplished with certification, although recent development of density-functional-theory (DFT) calculations help with the identification.\cite{Beeler90,Estreicher05,Ehlers06,Sharan17,Vincent20}
The high reactivity of Cu atoms with other defects,\cite{Keller90,Mesli92,Aboelfotoh95,West03,Latham05} whether intentionally introduced or not, make the identification difficult. 

Among the proposed structures of Cu defects, one of particular interest is the Cu$_{4}$ complex.
The advent of ultrahigh-resolution photoluminescence spectroscopy (UHR-PL) has greatly improved our understanding of defects in semiconductors.\cite{Cardona05,Steger11}
In the PL spectrum of Cu-containing Si, there is a sharp band at 1.014 eV, which corresponds to a zero-phonon (ZP) emission. The corresponding defect is called the Cu$_{\rm PL}$ center.
For a long time, this center was believed to be due to a Cu pair,\cite{Weber82} presumably a Cu$_{(s)}$-Cu$_{(i)}$ pair.\cite{Istratov98,Estreicher03,Estreicher05} A uniaxial-stress experiment showed that the site symmetry of the Cu$_{\rm PL}$ center is $C_{3v}$ symmetry.\cite{Weber82}
By applying UHR-PL to isotope-enriched Cu-containing silicon, it was revealed that the structure of the Cu$_{\rm PL}$ center is Cu$_{4}$: the isotope splitting of the Cu$_{\rm PL}$ center into five peaks is indisputable evidence of the involvement of four Cu atoms.\cite{Thewalt07,Steger08} 
Later, DFT calculations confirmed that the structure is composed of one substitutional Cu and three interstitial Cu at tetrahedral sites, Cu$_{(s)}$Cu$_{3(i)}$.\cite{Shirai09,Carvalho11} Here, we call this structure $C$ type after its $C_{3v}$ symmetry (see the left side of Fig.~\ref{fig:struct}).
A systematic study by Carvalho {\it et al.}\cite{Carvalho11} convinced researchers of this structural model of $C$-type. The agreement based on the local $C_{3v}$ symmetry is particularly convincing.

Since then, the formation mechanism of Cu$_{\rm PL}$ has been reappraised with the new understanding of the defect structure. Nakamura and coworkers, examined the depth profile of the Cu$_{\rm PL}$ center, and constructed the formation mechanism.\cite{Nakamura09,Nakamura10,Nakamura12} 
Yarykin and coworkers studied the dissociation of Cu$_{\rm PL}$ in a hydrogenation process.\cite{Yarykin13,Yarykin13a,Yarykin14} The dissociation of Cu$_{\rm PL}$ by hydrogenation has also been studied theoretically by the present authors (presented at 29th Int.~Conf.~Defects in Semiconductors, Matsue, Japan, 2017). 
During this study, we found a new type of the Cu$_{4}$ complex, which is more stable than $C$-type. In this paper, we call this $T$-type after its $T_{d}$ symmetry (see the right side of Fig.~\ref{fig:struct}). We soon became aware that this type is the same as that discovered by Tarnow for Li-doped silicon.\cite{Tarnow92} Aiming at the termination of a Si vacancy, Tarnow studied the effect of Li doping and found that the arrangement of four interstitial Li atoms around a vacancy is the lowest-energy state. This complex is called Li-saturated vacancy (Li$_{4}V$). He found that the formation of this type is a common feature for column I and VII dopants. This feature is reasonably expected for column IB dopants too. We have found that this is indeed the case for Cu. Consistent with this new view, Sharan {\it et al.}~pointed out that $C$-type of Cu$_{4}$ can be ruled out for the explanation of the Cu$_{\rm PL}$ emission on the basis of their calculation of the gap states.\cite{Sharan17}

In this paper, we present a theoretical study on the structure and electronic properties of the $T$-type Cu$_{4}$ complex. Although the basic structural model has already been proposed by Tarnow, in view of the importance of the Cu$_{\rm PL}$ center, we have investigated the new type of Cu$_{4}$ complex in terms of its detailed structural and electronic properties. Although the role of dopant is the same as other column I and VII elements, we found an interesting difference in the bonding nature.
The paper is organized as follows. In Sec.~II, the calculation method along with the terminology used in this paper are given. Sections III and IV are the main part of the study and cover the atomic structure, electronic structure, and phonon properties. In Sec.~V, we discuss the inconsistency in the symmetry. Finally, the present results are summarized in Sec.~VI.

\section{Methods and definitions}
\label{sec:method}

The electronic structures of Cu-containing silicon were studied by DFT calculation using a pseudopotential method. In an early stage of our study, we examined Cu$_{n}$-H$_{m}$ complexes with a wide range of compositions and atomic configurations for the reason described in Sec.~\ref{sec:intro}. Most of the calculations were performed by the pseudopotential code Osaka2k \cite{Osaka2k}. The norm-conserved (NC) pseudopotentials of Troullier--Martins type \cite{TM91} were used for the ionic potential, and the LDA of the Perdew--Zunger form \cite{Perdew81} and the GGA of the Perdew--Burke--Ernzerhof form \cite{PBE96} were used for the electron-correlation potential. 
The kinetic cutoff energy for the plane-wave expansion was 70 Ry, and two-point sampling, $\Gamma$ and $R$, was used to calculate the total energy.  
All the stable structures of Cu$_{n}$-H$_{m}$ complexes were found by this code. However, in contemporary calculations for transition elements, NC-type is rarely used. Accordingly, the total energy of the obtained structures was recalculated by ultrasoft (US) pseudopotentials of the Vanderbilt type, \cite{Vanderbilt90} which are implemented in Quantum Espresso \cite{QE2009}. The same $k$-sampling was used. The cutoff energy was 30 Ry, from which a convergence of 0.1 eV was obtained for the formation energy of Cu$_{(i)}$. The values of the total energy shown in this paper are those obtained by the US method, unless otherwise stated.

Supercells of the conventional unit cell of size $2 \times 2 \times 2$ were used, which may be sufficient for the calculation of the total energy. The convergence with respect to the size of supercells was tested (see S1 of Supplemental materials), which confirms the adequacy of this treatment. However, calculations of the gap states using this size have artificial dispersions, yielding broadening in the gap states of about 0.2 eV. This influence of supercells must be kept in mind for interpreting the results.
For the DOS calculation and phonon calculation, the results obtained by Osaka2k code are shown in this paper.

Before proceeding to the main part of this study, the terminology employed in this paper is summarized, because different names for the same defects are often used in the literature.
Cu$_{(s)}$ has already been introduced as denoting the substitutional site ($S$ site). 
Although there are many interstitial sites, here only the tetrahedral interstitial site ($T$ interstitial), which is the lowest-energy state for a single Cu atom, is relevant, and hence Cu$_{(i)}$ is referred to as this site. On the line extending a Si$-$Si bond, there is an antibonding ($AB$) site behind the bond, whose location has a distance from the nearest Si atom of approximately half the bond length. The position of atom 57 in Fig.~\ref{fig:struct} is a representative of this site.

The generic name Cu$_{4}$ is used when there is no need to specify Cu positions.
More specifically, the notation Cu$_{(s)}$Cu$_{3(i)}$ is used to refer to $C$-type. Cu$_{4}V$ represents $T$-type, following the notation of Tarnow.\cite{Tarnow92} However, note that $T$-type is also created by displacing the Cu$_{(s)}$ atom in $C$-type toward an $AB$ site, as shown in Fig.~\ref{fig:struct}.
This means that $T$-type can be formed from both Cu$_{(s)}$ and $V$.
Another type of four-member complex is formed by the aggregation of four Cu$_{(i)}$ atoms around a Si atom. This is denoted as Cu$_{4(i)}$. This type, called $I$-type, is not discussed so much in this paper. 
A Cu$_{4}$ tetrahedron refers to the Cu$_{4}$ unit of $T$-type. This unit can also be looked upon as a dodecahedral Cu$_{4}$Si$_{4}$ unit by combining it with the surrounding Si$_{4}$ tetrahedron, as shown on the right-hand side of Fig.~\ref{fig:struct}.

Experimentally, the defect states of Cu-related centers are detected by various methods. In most cases, the attribution to concrete defect structures is not certain.
Three types of defect states have been identified by DLTS measurement: \cite{Brotherton87} a Cu$_{\rm DLA}$ center at $E_{v}+0.23$, $E_{v}+0.42$, and probably a $E_{c}-0.16$ eV; a Cu$_{\rm DLB}$ center at $E_{v}+0.09$ eV; a Cu$_{\rm DLC}$ center at $E_{v}+0.40$ eV. By examining the correlations between DLTS and PL spectra, it was clarified that Cu$_{\rm DLB}$ center has the same origin as the Cu$_{\rm PL}$ center.\cite{Weber82,Brotherton87,Erzgraber95}
A DFT study by Sharan {\it et al.}~provides a useful guide for the identification of Cu-related centers. \cite{Sharan17}

\section{Atomic structure of Cu$_{4}$}
\subsection{Cu$_{4}$ tetrahedron} 
\label{sec:structure}

\begin{figure}[htbp]
\centering
\includegraphics[width=7.0 cm, bb=0 0 226 212]{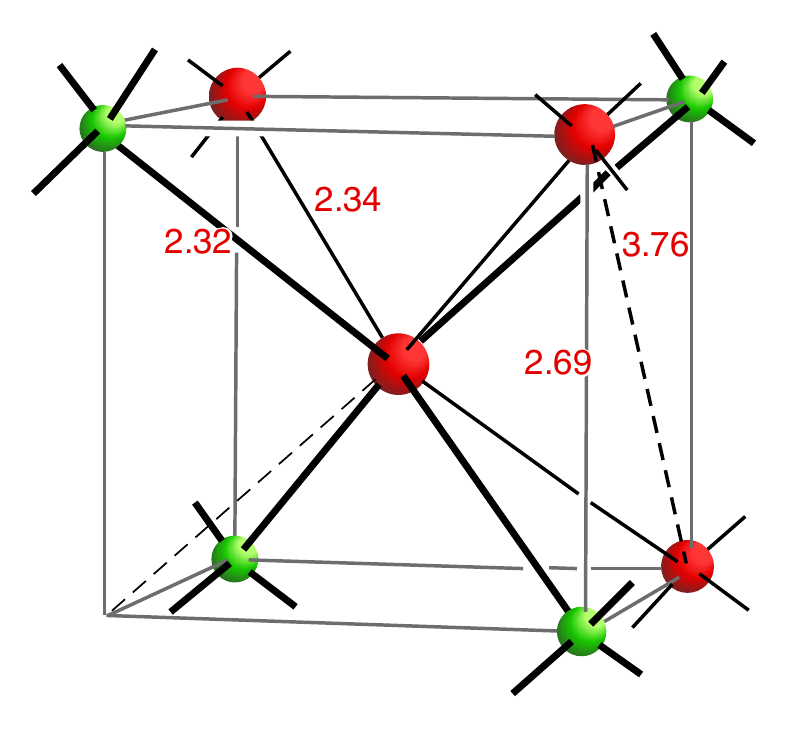}
\hspace{0.6 cm}
\includegraphics[width=7.0 cm, bb=0 0 226 212]{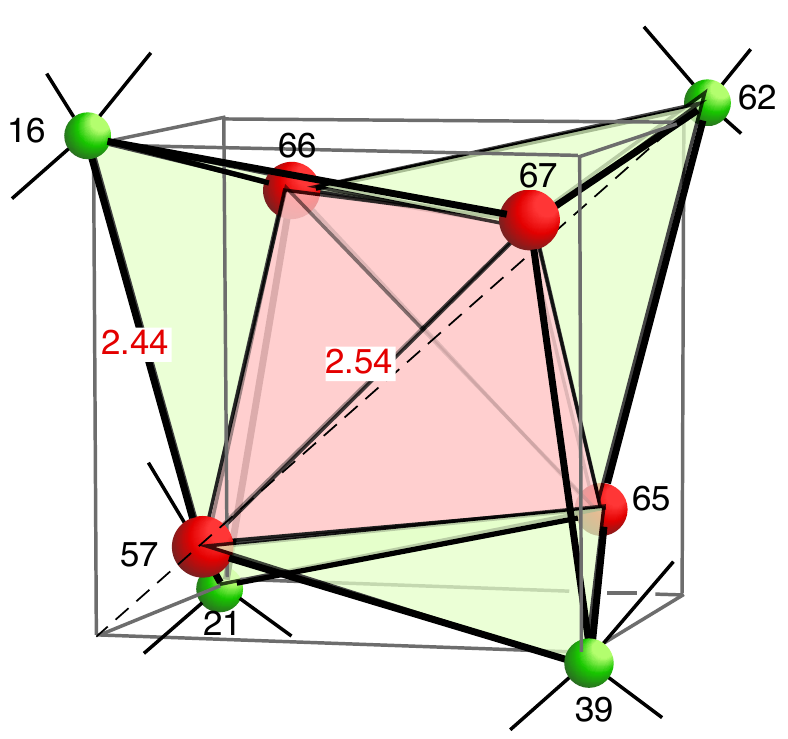}
\caption{
Comparison of $C$- (left) and $T$-type (right) Cu$_{4}$ complexes. Red spheres denote Cu and green spheres denote Si. In $C$-type, while the local symmetry of the substitutional Cu$_{(s)}$ is lowered to $C_{3v}$ about the $(111)$ direction, the displacements of the adjacent atoms around Cu$_{(s)}$ are very small. 
In $T$-type, Cu$_{(s)}$ moves toward an $AB$ site and drags three interstitial Cu$_{(i)}$ atoms with it, forming a tetrahedron. This Cu tetrahedron, together with the nearest Si tetrahedron, forms a  Cu$_{4}$Si$_{4}$ dodecahedron. The bond lengths  in \AA \  are indicated by red numbers unit. Serial numbers are attached to the atoms to label individual atoms. 
} \label{fig:struct}
\end{figure}

Figure~\ref{fig:struct} shows a comparison in the structure of Cu$_{4}$ complexes between $C$- and $T$-types. Previously, only $C$-type was known. For $C$-type, the substitutional Cu$_{(s)}$ is not displaced from the equilibrium position of the host crystal. Other three Cu$_{(i)}$ atoms are located at interstitial $T$ sites, with each being in a direction of $AB$ site. Accordingly, the bond length of ${\rm Cu_{(s)}-Cu_{(i)}}$ is almost the same as that of the host Si$-$Si bond (2.35 \AA). The local symmetry of Cu$_{(s)}$Cu$_{3(i)}$ is $C_{3v}$.
When the Cu$_{(s)}$ in $C$-type is displaced toward an $AB$ site, for example, in the $(111)$ direction in Fig.~\ref{fig:struct}, $T$-type is formed.  In $T$-type, the Cu$_{(s)}$ is located close to an $AB$ site. This movement of the Cu$_{(s)}$ drags the three Cu$_{(i)}$ atoms a long distance. While the bond ${\rm Cu_{(s)}-Cu_{(i)}}$ is elongated to 2.54 \AA, the interatomic distance between two Cu$_{(i)}$ atoms is markedly reduced from 3.57 to 2.54 \AA. For each triangular surface of the Cu tetrahedron, the nearest neighboring Si atom is located at the vertex of a triangular pyramid. As a whole, four Cu atoms together with four Si atoms form a dodecahedral complex. The local symmetry is recovered to the original $T_{d}$ symmetry. By changing from the $C$-type to $T$-type configuration, the energy gain is 0.28 eV.

It is not uncommon to find the four-atom aggregation of foreign atoms in covalent materials. For example, in diamond crystal, four N atoms gather around a vacancy and form a so-called B center (${\rm N_{4}}V$).\cite{Diamond-EMIS,Mainwood94} However, in this case, four N atoms are incorporated in substitutional sites, and thereby this center is appropriately expressed as N$_{4(s)}V$. There are no polyhedral units. In the present case of $T$-type, the coordination number of the Si atoms in the dodecahedron becomes six; among these coordinations, three are used to bond Cu atoms. Si clathrate compounds\cite{Yamanaka10} and fullerenes\cite{Fullerenes94} have polyhedral units. However, the coordination number of the atoms in these crystals is still as small as four, which reflects the covalent character. Structural units of polyhedra are found in metals and quasicrystals, which have higher coordination numbers. It is indeed rare to find such polyhedra in tetrahedrally coordinated materials. 
In the past, a few cases that indicate polyhedral units in the tetrahedrally coordinated network were discovered. Heavily B-doped silicon ($> 10^{20}\ {\rm cm}^{-3}$) with an agglomeration of B atoms forming an icosahedral B$_{12}$ unit was reported.\cite{Mizushima94,Yamauchi97} However, in the present case, the concentration of Cu is very low, possibly much less than $10^{16}\ {\rm cm}^{-3}$.

\begin{table}
\caption{
Formation energy $\varepsilon_{f}$ and binding energy $\varepsilon_{b}$ of Cu$_{n}$ complexes calculated by US potentials. For the group denoted as Cu$_{n(i)}$, the complexes are composed of Cu$_{(i)}$ atoms only. For the group denoted as Cu$_{(s)}$Cu$_{n-1(i)}$, $n-1$ atoms of Cu$_{(i)}$ are aggragated around a Cu$_{(s)}$ atom. 
Numbers in parentheses indicate the values obtained by NC potentials. All the energies are given in eV.
}
\begin{ruledtabular}
\begin{tabular}{c| ccc|cccc}
  \multicolumn{1}{c|}{ $n$} & \multicolumn{3}{c|}{Cu$_{n(i)}$} & \multicolumn{4}{c}{Cu$_{(s)}$Cu$_{n-1(i)}$}    \\ 
   \cline{2-8}
   & $ \varepsilon_{f} $ & & $\varepsilon_{b}$ & & $ \varepsilon_{f} $ & & $\varepsilon_{b}$  \\ 
  \hline
  1  & 1.27 & (2.05) & & & 2.02 & (2.01) &  \\ 
  2 & 1.25 & (1.99) & 0.12 & & 1.27 & (1.42) & 0.84    \\
  3 & 1.24 & (1.98) & 0.15 & & 1.04 & (1.27) & 0.78   \\ \hline
  4 & 1.22 & (1.92) & 0.18 & $C$-type & 0.95 & (1.17) & 0.69  \\
     &      &      & & $T$-type & 0.88 & (1.03) & 0.96  \\
\end{tabular}
\end{ruledtabular}
\label{tab:DEvsCun}
\end{table}%

Table \ref{tab:DEvsCun} shows a comparison of the formation energy and binding energy among various Cu$_{n}$ complexes. The formation energy $E_{f}[\textrm{Cu$_{n}$}]$ of a complex Cu$_{n}$ in silicon is defined by taking the reference to the bulk Si and metal Cu as
\begin{equation}
E_{f}[\textrm{Cu$_{n}$}] = E[\textrm{Cu$_{n}$:Si}] - \left\{ nE[{\rm Cu^{(\rm bulk)}}] + E[{\rm Si}] \right\},
\label{eq:def-Ef}
\end{equation}
where $E[{\rm A}]$ is the total energy of system A. We use a notation $E[{\rm A}]$ in place of the usual notation $E_{\rm A}$ because the only information that we are interested in is the name of material A. 
In the table, the formation energy is given by the energy per Cu atom, $\varepsilon_{f}[\textrm{Cu$_{n}$}] = E_{f}[\textrm{Cu$_{n}$}]/n$. The binding energy $\varepsilon_{b}$ of Cu$_{n}$ is defined as the energy gain upon bringing a Cu$_{(i)}$ atom into the preexisting complex Cu$_{n-1}$,
\begin{equation}
\varepsilon_{b}[\textrm{Cu$_{n}$}] = E[\textrm{Cu$_{n}$:Si}] - \left\{ E[\textrm{Cu$_{n-1}$:Si}] + E[\textrm{Cu}_{(i)}:\textrm{Si} ] \right\}.
\label{eq:def-Eb}
\end{equation}
In the table, the first group, denoted by Cu$_{n(i)}$, indicates complexes composed of $n$ Cu$_{(i)}$ atoms around a Si atom. The second group, denoted by Cu$_{(s)}$Cu$_{n-1(i)}$, indicates complexes of one Cu$_{(s)}$ surrounded by $(n-1)$ Cu$_{(i)}$ atoms. $T$-type Cu$_{4}$ is here classified as a member of Cu$_{(s)}$Cu$_{n-1(i)}$, even though there is no Cu$_{(s)}$ atom. 
Although the general trend with changing $n$ agrees between the US and NC potentials, there is a systematic difference between them in that the values of $\varepsilon_{f}$ for Cu$_{n(i)}$ calculated by NC potentials are larger than those calculated by US potentials. For $n=1$, the calculation by NC potentials results in $\varepsilon_{f}[{\rm Cu}_{(i)}] > \varepsilon_{f}[{\rm Cu}_{(s)}]$. Generally, it is believed that $\varepsilon_{f}[{\rm Cu}_{(i)}]$ is lower than $\varepsilon_{f}[{\rm Cu}_{(s)}]$, for example, by 0.7 eV by Latham {\it et al.}\cite{Latham05} Accordingly, the results calculated by US potentials are employed in the following. We suppose that the used NC potential for Cu is less extended compared with the US potential.

For the group Cu$_{n(i)}$, the formation energy per atom $\varepsilon_{f}$ is almost the same irrespective of $n$. This implies that there is virtually no binding among interstitial Cu$_{(i)}$ atoms. In contrast, there is a clear decrease in $\varepsilon_{f}$ for the group Cu$_{(s)}$Cu$_{n-1(i)}$. This means that a Cu$_{(i)}$ atom mostly interacts with Cu$_{(s)}$ only. $T$-type Cu$_{(s)}$Cu$_{3(i)}$ has the strongest binding energy. This has already been shown by Carvalho {\it et al.} \cite{Carvalho11} From this observation, it follows that a substitutional Cu$_{(s)}$ can be a seed for forming $T$-type Cu$_{(s)}$Cu$_{3(i)}$. $T$-type can, however, be also seen as an aggregate of four Cu$_{(i)}$ around a vacancy, which is evident from Fig.~\ref{fig:struct}. Thus, both Cu$_{(s)}$ and a vacancy can be a seed for $T$-type Cu$_{4}$.

The tendency that the Cu$_{(s)}$ in Cu$_{(s)}$Cu$_{n-1(i)}$ is increasingly likely to be deflected from the substitutional site as $n$ increases is observed by comparing the difference in $\varepsilon_{f}$ when the Cu is placed at an $S$ site and at an $AB$ site. For $n=1$, the $\varepsilon_{f}$ of Cu$_{(s)}$ located at an $AB$ site is higher than that at an $S$ site by 0.8 eV; an AB site is a location of only local energy minimum. This energy difference decreases as the number $n$ in Cu$_{(s)}$Cu$_{n-1(i)}$ increases, and finally the order is reversed at $n=4$.
The increasing favorarability of forming the tetrahedron complex can be interpreted as the change in the bonding character. In the original Si crystal, Si$-$Si bonds have a strong covalent character, which possesses highly oriented distribution of the valence electrons. This situation is suitable for a Cu$_{(s)}$ point defect. As the number of Cu atoms around the center Cu$_{(s)}$ is increased, the valence electrons becomes increasing concentrated around the center and the orientation dependence of the charge distribution is weakened, suppressing the restoring force against deviation from the $S$ site. When $n=4$, that is, $C$-type Cu$_{(s)}$Cu$_{3(i)}$, the coordination number of the center Cu$_{(s)}$ becomes six. This coordination numbers of six is a common property when a crystal behaves as a metal. Interestingly, the Cu$-$Cu bond length in the Cu tetrahedron is 2.54 \AA, which is very close to that of metallic Cu (2.55 \AA). The Cu atoms in the Cu tetrahedron retain their natural bond length in the bulk. 
This matching of the bond length with the bulk value is a special case for Cu only. For other isovalent complexes of $M_{4}$ type ($M$=H, Li, and Br), the $M-M$ bond length is different from its bulk value. See Supplemental materials (S2).

\subsection{Energy landscape}
\label{sec:energetics}

\begin{figure}[htbp]
\centering
\includegraphics[width=65 mm,bb=0 0 595 765]{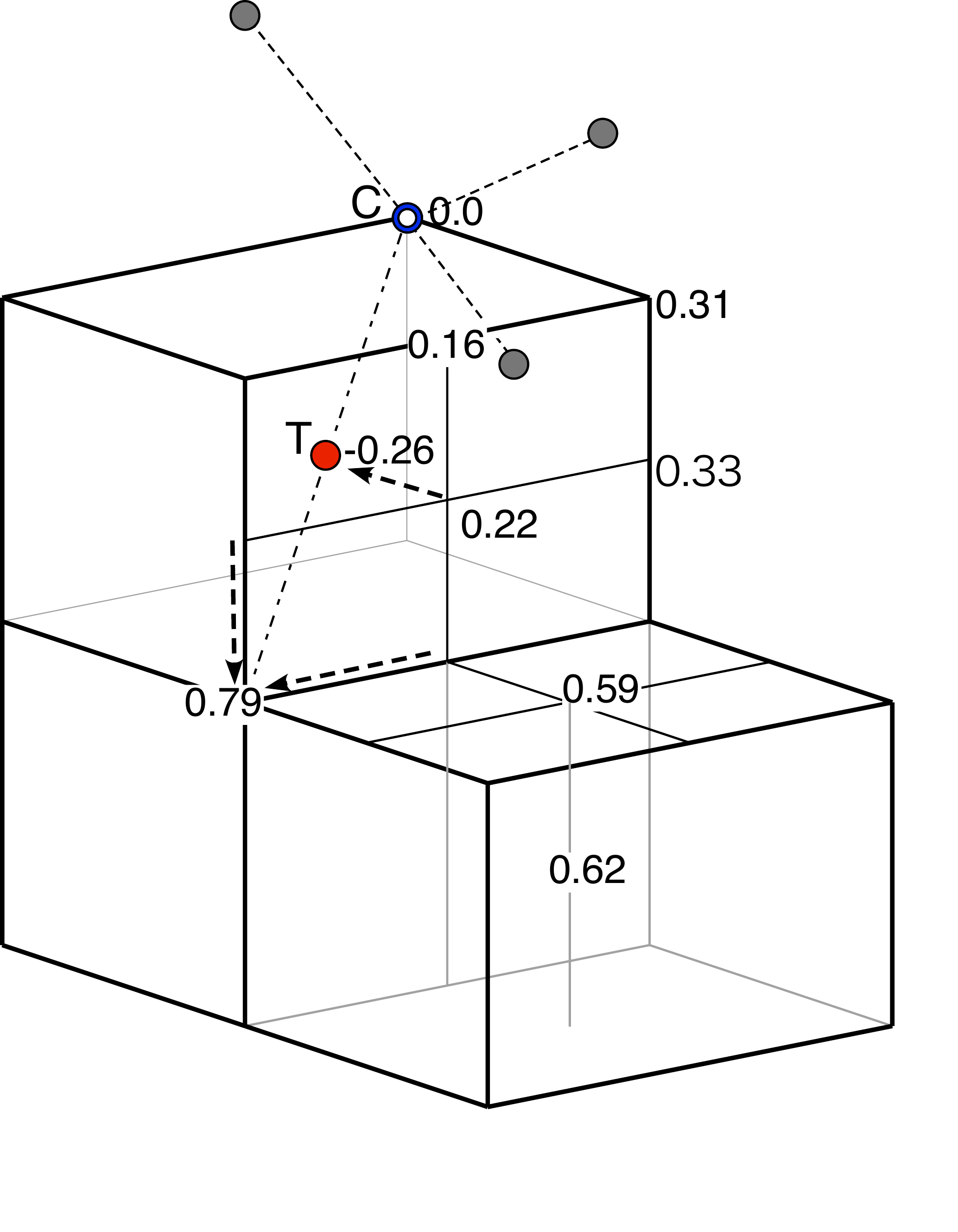}

\caption{
Energy landscape of Cu$_{4}$ calculated by US potentials. 
The initial configuration of $C$-type, Cu$_{(s)}$Cu$_{3(i)}$, is shown, with gray spheres indicating the three Cu$_{(i)}$ atoms. 
By changing the starting position of Cu$_{(s)}$, the local energy-minimum structures were calculated. Unnormalized formation energy $E_{f}$ is indicated at each starting position of Cu$_{(i)}$. For the case of large displacement, the displacement is shown by an arrow. The origin of $E_{f}$ is taken to be that of $C$ type. 
} \label{fig:map-cu4}
\end{figure}

To confirm that the $T$ site is the globally energy-minimum position of Cu$_{4}$, we calculated the energy landscape of Cu$_{4}$. Starting from $C$-type Cu$_{(s)}$Cu$_{3(i)}$, we constructed the initial structures by changing the position Cu$_{(s)}$ in Cu$_{(s)}$Cu$_{3(i)}$ and then optimized the structures. Although the name Cu$_{(s)}$ has its exact meaning only when at an $S$ site, we continue to use this name here. The results are shown in Fig.~\ref{fig:map-cu4}. In all the cases, the formation energy $E_{f}$ was higher than that of $T$-type. In some cases, Cu$_{(s)}$ spontaneously moved to the $AB$ site. Furthermore, no case was found in which $E_{f}$ was lower than even that of $C$ type, when Cu$_{(s)}$ was located far from the $AB$ site. $C$ type can exist as a metastable state.

\begin{figure}[htbp]
\centering
\includegraphics[width=11.0 cm, bb=0 0 370 260]{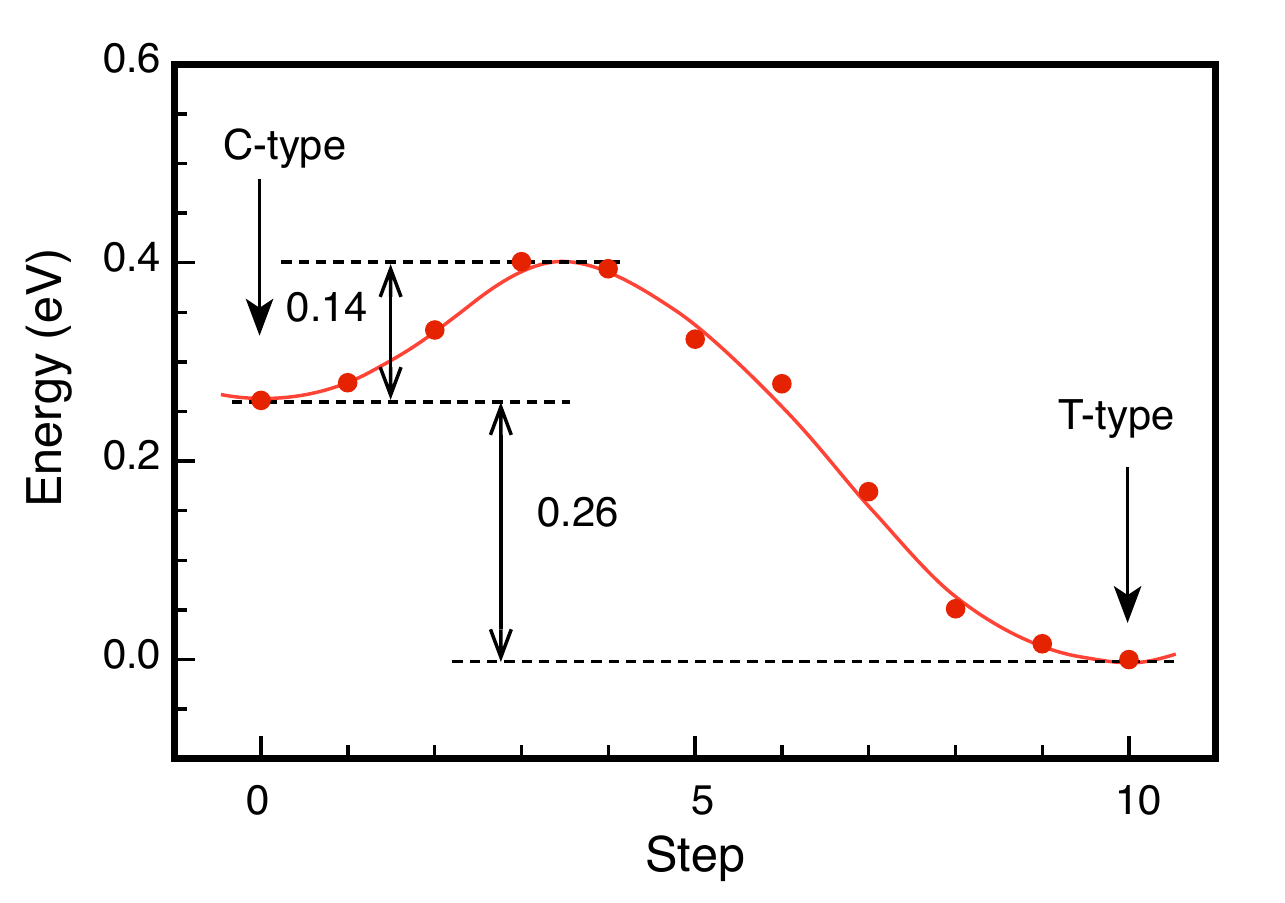} \\
\caption{
Adiabatic potential between $C$- and $T$-type structures calculated by US potentials. The total energy relative to that of $T$-type is plotted with the whole path into 10 steps. The abscissa approximately shows the relative position of the Cu$_{(s)}$ from the $S$ site to the $AB$ site.
} \label{fig:adiabatic}
\end{figure}

The adiabatic potential between $C$- and $T$-type structures was calculated by the nudged elastic band (NEB) method \cite{Henkelman00} and its variant implemented in Osaka2k code.
Figure \ref{fig:adiabatic} shows the adiabatic potential of Cu$_{(s)}$ in the Cu$_{4}$ complex along the direction from the $S$ to the $AB$ site, for example, the $(111)$ direction in Fig.~\ref{fig:struct}. The terminal positions in this figure correspond to $C$- and $T$-type structures.
The energy at the $AB$ site is 0.26 eV lower than that of the $S$ site (0.51 eV calculated by NC potentials). There is an energy barrier of 0.14 eV from the $S$ to the $AB$ site (0.08 eV calculated by NC pseudopotentials). 
This barrier is small compared with that of the reverse reaction. If the Cu$_{4}$ complex is formed in $C$-type, annealing at several hundred $^{\circ}$C will eliminate this type of structure, converting it to $T$-type.
From all the above results, we conclude that the $T$-type Cu$_{4}$ complex is the most stable structure.

\subsection{Symmetry inconsistency}
\label{sec:symmetry-inconsistency}
The most interesting question in the Cu$_{4}$ problem is which structure, $C$-type or $T$-type, is the cause of the Cu$_{\rm PL}$ band. From the above energetic consideration, $T$-type must be the Cu$_{\rm PL}$ center. However, this conclusion suffers from a serious weakness regarding symmetry. 
The experiment of the uniaxial stress on the PL spectroscopy unambiguously indicates that the symmetry of the Cu$_{\rm PL}$ center is $C_{3v}$.\cite{Weber82} Several scenarios have been investigated in an attempt to resolve this inconsistency. 

(1) The observed Cu$_{\rm PL}$ emission stems from a charge state of Cu$_{4}$. By changing the charge state, the atom positions can be displaced. Our calculation shows that the Cu$_{4}$ tetrahedron is indeed deformed to the $C_{3v}$ pyramid by changing the charge state (see Table IV Supplemental materials). 
The symmetry inconsistency thus could be explained that the observed PL emission stems from the charge state of Cu$_{4}$. The problem of this scenarios is that our calculation shows that charge states of Cu$_{4}$ appear only for $p$-type silicon. For $n$-type silicon, Cu$_{4}$ must be neutral. However, experiments show that the Cu$_{\rm PL}$ emission is observed for both $p$- and $n$-type wafers.\cite{NakamuraCommun}

(2) Metastable $C$-type is the true origin of the Cu$_{\rm PL}$ emission. $C$-type can exist under suitable conditions, even though it is metastable. In this case, the symmetry inconsistency does not occur. In addition, this scenario appears to explain the dependence of the concentration of the Cu$_{\rm PL}$ centers on the growth temperature. Generally, the Cu$_{\rm PL}$ emission is observed in samples prepared first by diffusion of Cu at high temperatures of around 800$^{\circ}$C and then by quenching to room temperature. The Cu$_{\rm PL}$ emission disappears upon annealing above 200$^{\circ}$C.\cite{Weber82,Nakamura09} From this observation, it is reasonable to regard the Cu$_{\rm PL}$ center as a metastable state at room temperature. 
Hence, elimination of the Cu$_{\rm PL}$ emission upon annealing at moderate temperatures is consistent with the metastability of $C$-type shown in this study. A problem of this scenario is that there is no reason why only $C$-type is PL active, and why $T$-type is not.

(3) Electronic excited states yield atom distortions from $T_{d}$ symmetry. 
What is actually observed in PL spectra is the excited state induced by light illumination.
The excited states $\Gamma_{3}$ and $\Gamma_{4}$, which exhibit an isotope shift, are degenerate states and therefore they induce a Jahn--Teller distortion if these states are partially occupied. This idea is the same as Tarnow's proposal for Li$_{4}V$.\cite{Tarnow92} We consider that this is the most likely scenario to occur. The complicated isotope splitting of Cu$_{\rm PL}$ can be interpreted on the basis of this scenario. Details of this mechanism will be presented elsewhere.

\section{Electronic structure}
\subsection{Distribution of charge density}
\label{sec:chds}

\begin{figure}[htbp]
\centering
\includegraphics[bb={0 0 360 259},width=9 cm]{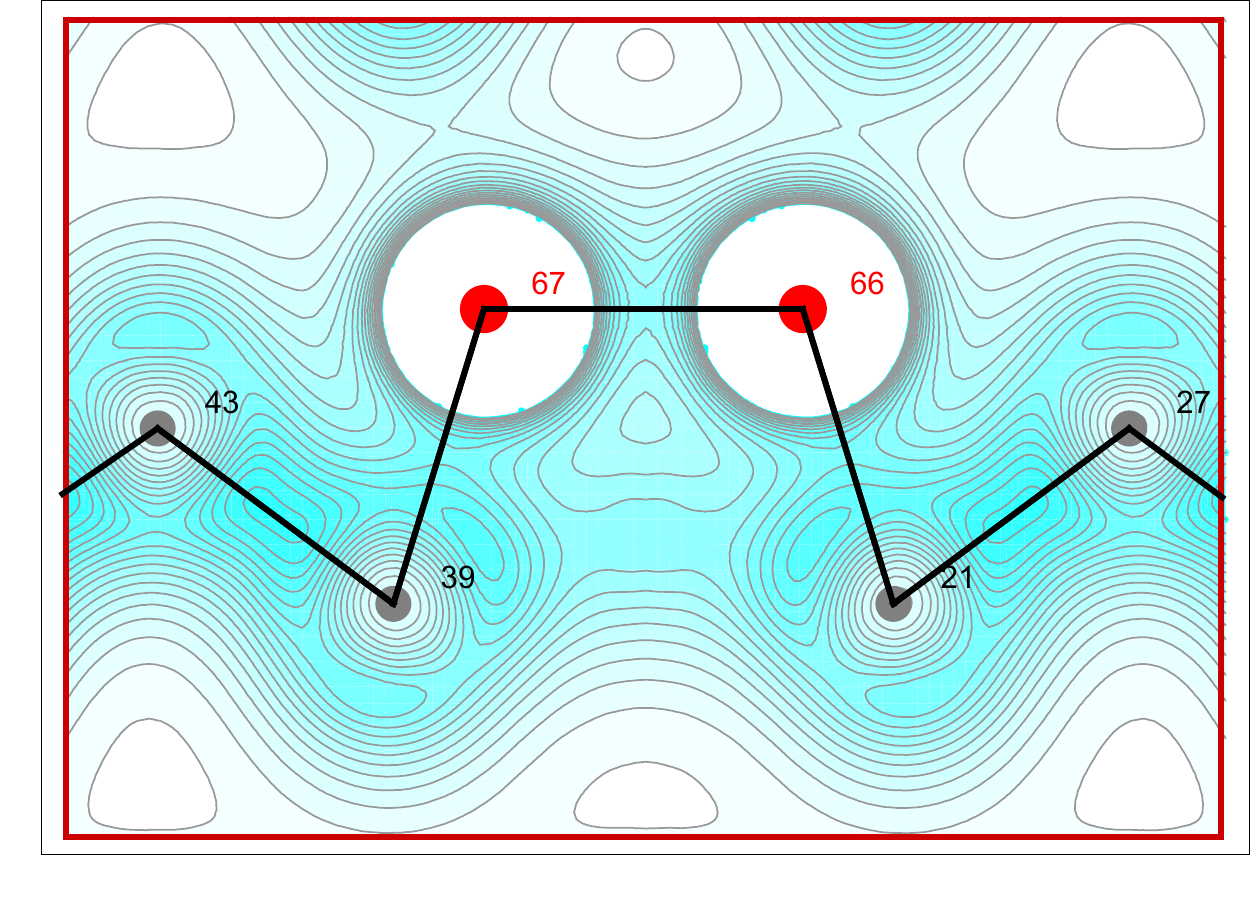} 
\vspace{0.5 cm}
\includegraphics[bb={0 0 360 188}, width=9 cm]{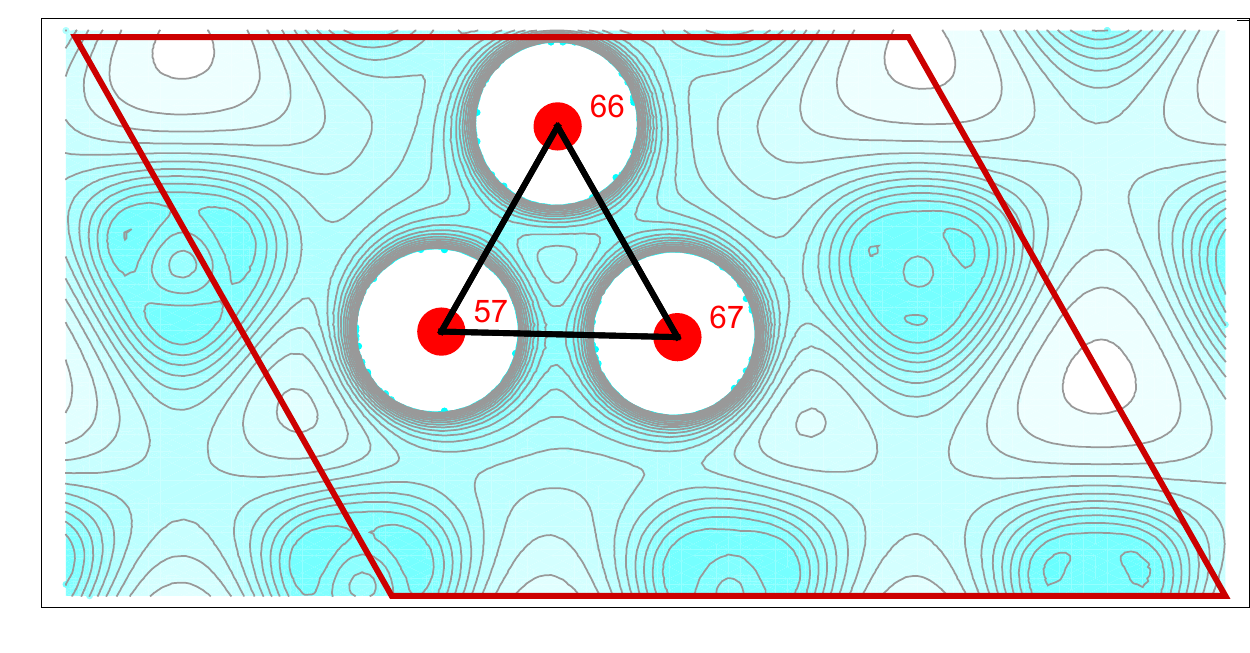}
\caption{Distribution of the charge density $\rho({\mathbf r})$ of $T$-type. The density is plotted in the range from 0 to 0.1 $el/a_{\rm B}^{3}$ in 20 steps. The atom labels are the same as in Fig.~\ref{fig:struct}. Red circles indicate Cu atoms. Since $\rho({\mathbf r})$ around Cu atoms exceeds the maximum value of the range, it is truncated.} \label{fig:rho1}
\end{figure}

\begin{figure}[htbp]
\centering
\includegraphics[bb={0 0 360 361},width=6.5 cm]{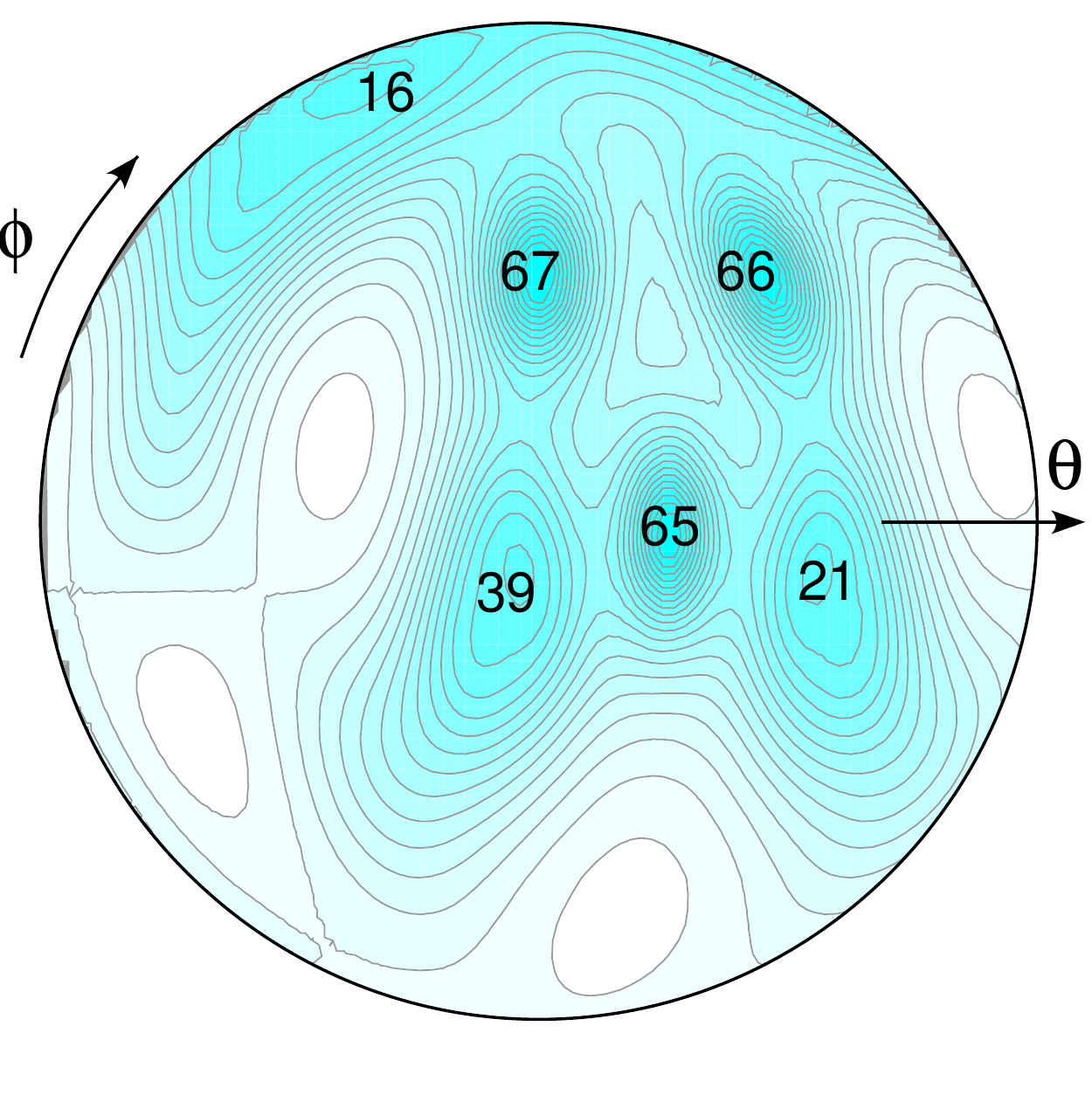} 
\caption{Distribution of the charge density of $T$-type in the spherical coordinates $(\theta, \phi)$ around a Cu 57 with a radius 1.6 \AA. The density is plotted with a maximum of 0.1 $el/a_{\rm B}^{3}$ in 20 steps. The direction of the view from the central atom 57 is indicated by the label of the atom which is seen in that direction. The atom labels are the same as in Fig.~\ref{fig:struct}. } \label{fig:rho2}
\end{figure}

The unusual structure of $T$-type Cu$_{4}$ requires a close inspection of the electronic structure.
Figures \ref{fig:rho1}  and \ref{fig:rho2} show the charge distribution $\rho({\mathbf r})$ of $T$-type Cu$_{4}$. The upper panel of Fig.~\ref{fig:rho1} shows $\rho({\mathbf r})$ in a cut plane on which the shortest Cu$-$Si bonds lie.
For the $39-67$ Cu$-$Si bond (see Fig.~\ref{fig:struct} for atom labels), the maximum $\rho_{\rm Cu-Si}$, which is found near the center, is 65 m$el/a_{\rm B}^{3}$. Compared with that of bulk Si, {\it i.e.}, 88 m$el/a_{\rm B}^{3}$, we see that the covalent character is weakened. This can be expected from the fact that the Cu$-$Si bond is elongated from the original Si$-$Si bond by 3.4\%.
An interesting fact of $T$-type is that the maximum value of $\rho_{\rm Cu-Si}$ is larger than that of $C$-type, 50 m$el/a_{\rm B}^{3}$, in spite of the fact that the bond length is elongated from that of $C$-type by 5\% (see Fig.~\ref{fig:struct}). This is due to an interference effect where each apex Si atom of the Cu$_{4}$Si$_{4}$ dodecahedron ties up three Cu$-$Si bonds, so that the charge density of one Cu$-$Si bond is enhanced by the other two bonds.

The lower panel of Fig.~\ref{fig:rho1} shows $\rho({\mathbf r})$ in a cut plane on which a Cu triangle lies. Despite the center of the Cu triangle being an interstitial site, a high density of electrons, 35 m$el/a_{\rm B}^{3}$, is observed. This value is comparable to the magnitude at the center of the Cu$-$Cu bond, 45 m$el/a_{\rm B}^{3}$, in the cut plane. This indicates that the dodecahedron has a relatively uniform charge distribution, which is a typical characteristic of metals. In fact, the fcc metal Cu has a uniform distribution of the electron density ranging from 20 to 40 m$el/a_{\rm B}^{3}$. The charge density of the Cu$_{4}$ tetrahedron falls in this range, implying metallic character.
A simple method of electron counting also supports this viewpoint. A Cu atom of the Cu$_{4}$ tetrahedron is shared with three triangles, and donate one electron to the three triangles. One triangle has 1/3 electron contributed from each of the three apex Cu atoms, resulting in a total of one electron. In this manner, a covalent bond in the Cu$_{4}$Si$_{4}$ dodecahedron is formed between the one end of Si atom and the other end of a triangular surface formed by Cu atoms. The latter end broadens the highly oriented character of the covalent bond possessed by the original Si$-$Si bond.

An unusual characteristic of the Cu bond in the dodecahedron can be seen in a spherical map of $\rho({\mathbf r})$  around a Cu atom. Figure \ref{fig:rho2} shows a plot of $\rho({\mathbf r})$ on a spherical surface $(\theta, \phi)$ around Cu atom 57. The spherical surface is cut at a distance of 1.6 \AA\ from atom 57. We see that the charge density is concentrated in approximately half the entire of the solid angle, as is evident from the geometry of the dodecahedron. In many respects, the Cu$_{4}$ tetrahedron has the appearance of a metallic unit.
This is in contrast to the case of the $T$-type Li$_{4}$ complex, even though the same dodecahedron  is formed. Most of the electrons donated from Li atoms move to the tetrahedral bonds of the host network. This is reflected in the fact that the bond length of Li$-$Li in the Li$_{4}$ complex, 2.57 \AA, is considerably different from that of bulk Li, 3.04 \AA. See Supplemental materials (S2). This implies that the orbital character of the bulk Li is not an important factor in formation of the Li complex in silicon. The role of the Li atom is simply the termination of a dangling bond.


\subsection{DOS structures}
\label{sec:DOS}
\begin{figure}[htbp]
\centering
\includegraphics[width=10.0 cm]{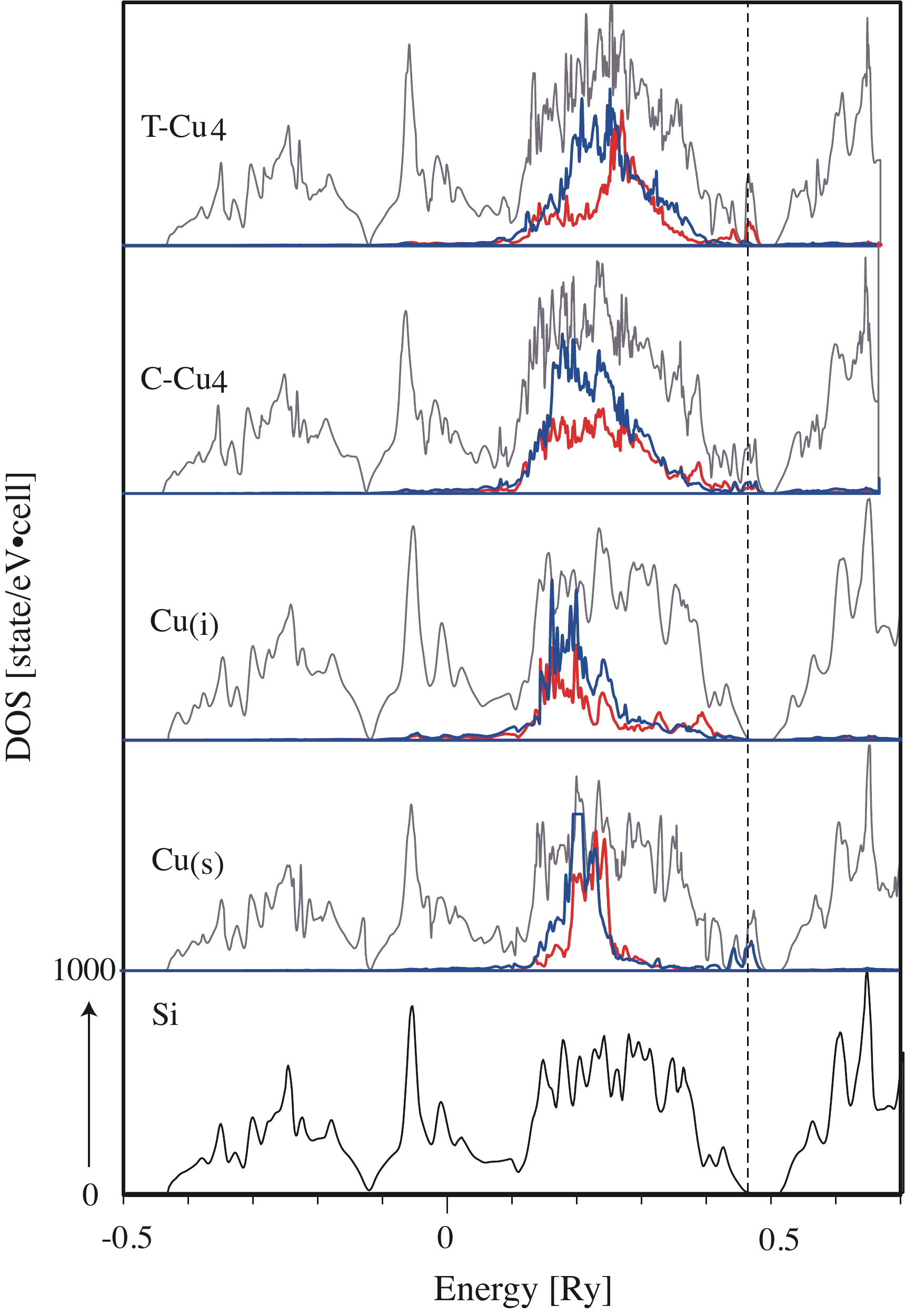} \\
\caption{
DOS spectra of $T$- and $C$-type Cu$_{4}$, compared with those of point defects Cu$_{(s)}$ and Cu$_{(i)}$ (gray lines). The projection to the atomic orbitals $d\gamma$ (red) and $d\varepsilon$ (blue) of Cu atoms is shown with a magnified scale. The top of the valence band is indicated by the dashed line.
} \label{fig:dos}
\end{figure}

\begin{figure}[htbp]
\centering
\includegraphics[width=8.0 cm]{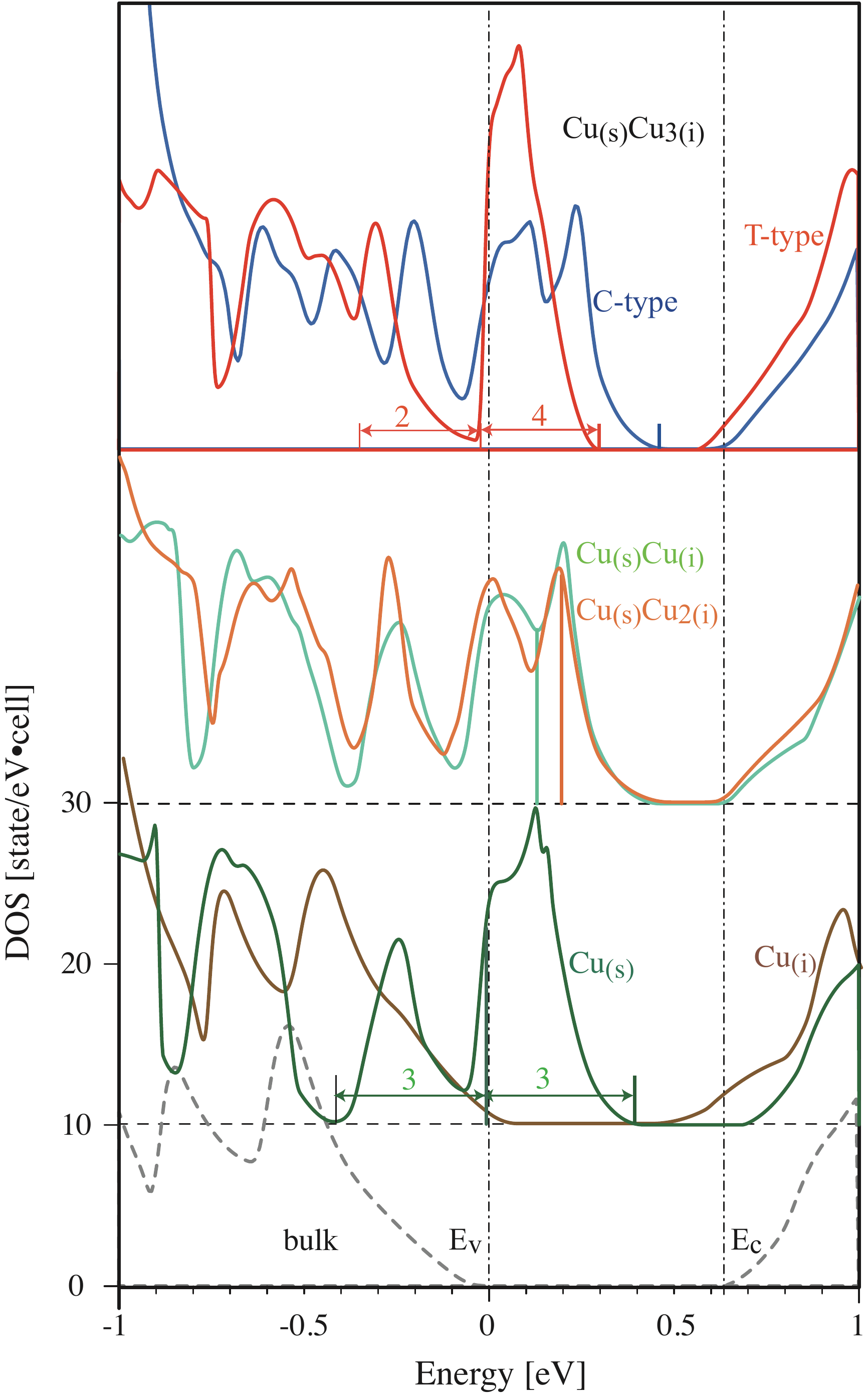} 
\caption{
DOS spectra around the band gap for Cu complexes. The top of the valence band is taken as the energy origin. Fermi levels are indicated by thick vertical lines. The arrows with numbers indicate the energy range and the number of electrons within the range.
} \label{fig:gap-dos}
\end{figure}

Let us next investigate the DOS structure of Cu complexes.
Figure \ref{fig:dos} shows a comparison of the DOS between $C$-type and $T$-type, together with those of point defects Cu$_{(s)}$ and Cu$_{(i)}$. In this figure, the projected DOS on Cu $d$ orbitals is also shown. Enlargements of the DOS spectra near the band gap in eV is shown in Fig.~\ref{fig:gap-dos}. The complex types Cu$_{2}$ and Cu$_{3}$ are included here for comparison.
In determining the gap structures among different systems, there is a technical problem of how to determine the energy origin among systems having different numbers of electrons. 
This problem cannot be overlooked, because of the fact that up to four Cu atoms are inserted in a relatively small supercell of size $2 \times 2 \times 2$. 
Various methods of correcting artifacts in use of supercells have been proposed.\cite{Makov95,Wright06a} Here, a simple method, which we believe to be physically sound, was employed. By comparing these DOSs over the entire energy range, we observed that the spectrum of the lower part of the valence band below -0.12 Ry, which is the so-called $s$ band, was the same for all the cases examined. Accordingly, the energy origin was adjusted so as to match the spectrum of the $s$ band.
Another point to be noted is that Kohn--Sham (KS) levels are one-particle eigenvalues. Donor and acceptor levels are different quantities. These are ionization energies, which must be obtained by the difference in the total energy between the charged state and neutral state. Some of results are shown in Supplemental material (S6). As mentioned above, the correction is more needed as the charge of Cu$_{4}$ increases, while the correction becomes is less reliable.\cite{Freysoldt14} Here, we content ourselves with studying DOS spectra. Despite this, the highest-occupied KS level agrees with the ionization energy, provided the effect of atom relaxation is negligible. 


As can be seen in Fig.~\ref{fig:gap-dos}, gap states appear for the Cu$_{(s)}$ and Cu$_{(s)}$Cu$_{3(i)}$ complexes. The point defect Cu$_{(i)}$ and a complex Cu$_{4(i)}$ (not shown here) have no clear gap state, with smearing of the band edge being the sole effect. Although Cu$_{(i)}$ can be regarded as a donor in the sense that it provides one electron, no donor level is created in the band gap. This is because the effect of adding Cu$_{(i)}$ is creating new bands, which lie in the middle of the $p$ valence band, whereas the effect of Cu$_{(s)}$ is mixing of the $d$ orbitals with the preexisting $p$ bands so that antibonding states result in. Cu$_{(i)}$ is electrically inactive.\cite{LW62} The complex Cu$_{4(i)}$ is essentially the sum of four independent Cu$_{(i)}-$Si bonds, as described in Sec.~\ref{sec:structure}.

It is known that a point defect Cu$_{(s)}$ creates three $d\varepsilon$ states in the band gap of Si, which can accommodate six electrons.\cite{Fazzio85,Beeler90} The neutral state of Cu$_{(s)}$ fills half of the $d\varepsilon$ states. These occupied states merge with the top of the valence band, leaving three empty states in the gap. These empty states are progressively occupied by adding Cu$_{(i)}$ atoms around Cu$_{(s)}$. This can be seen in the evolution of the gap states of Cu$_{(s)}$Cu$_{n(i)}$ in Fig.~\ref{fig:gap-dos}. By forming $C$-type, Cu$_{(s)}$Cu$_{3(i)}$, the three gap states $d\varepsilon$ are completely occupied by three additional electrons supplied from three Cu$_{(i)}$ atoms.\cite{Carvalho11} 
The calculated KS levels of the gap states are 0.10 and 0.21 eV above the top of the valence band, with the latter corresponding to the main peak. Because of the fact that the width of band gap is generally underestimated in DFT calculations, the values of these gap states is too high to account for the energy of 1.014 eV of the Cu$_{\rm PL}$ band, which is the basis on which Sharan {\it et al}.~suspected the $C$-type as the origin of the Cu$_{\rm PL}$ band.\cite{Sharan17} 

The band filling scheme of $T$-type is the same as that of $C$-type, and $T$-type exhibits gap states just above the top of the valence band similarly to $C$-type. However, the peak position of the gap states is lowered to 0.1 eV. This KS level is coincident with the Cu$_{\rm DLB}$ level, which is considered to correspond to the 1.014-eV band of the Cu$_{\rm PL}$ center, although the exciton level is not necessarily the same as that of the defect level \cite{YuCardona}.
Thus, it is likely that $T$-type is responsible to the Cu$_{\rm PL}$ band. The highest density at this peak also accounts for the intense luminescence of Cu$_{\rm PL}$ center.
Further study of impurity levels as done by Sharan {\it et al}.~\cite{Sharan17} did is desirable.



The stability of $T$-type Cu$_{4}$ can be explained formally in accordance with this valence-filling scheme, as discussed above. All the gap states are fully occupied, similar to the case of $C$-type. However, the orbital character of the gap states is different. As can be seen in Fig.~\ref{fig:dos}, the gap states of $T$-type have $d\gamma$-orbital character. This molecular-orbital character can be seen more clearly by plotting wavefunctions, which are given in S3 of Supplemental materials. This is contrary to the case of Cu$_{(s)}$. For the reason of symmetry, only $d\varepsilon$ orbitals $\phi_{{\rm Cu}_{(s)}}^{\varepsilon}$ of Cu$_{(s)}$ can be coupled to the host valence band $\Psi_{\Gamma_{25'}}$ in a manifold of wavefunctions of the crystal $\{ \Psi_{k=0} \}$. The symmetry species $\Gamma_{25'}$ is equivalent to $t_{2}$ in Mulliken's notation for $d\varepsilon$ orbitals. The antibonding coupling between $\Psi_{\Gamma_{25'}}$ and $\phi_{{\rm Cu}_{(s)}}^{\varepsilon}$ creates a gap state $t_{2}$. 
For $T$-type, however, the molecular orbitals $\phi_{{\rm Cu}_{4}}^{t_{2}}$ of the Cu$_{4}$ tetrahedron that can be coupled to the host band $\Psi_{\Gamma_{25'}}$ are composed of different atom orbitals. These molecular orbitals $\phi_{{\rm Cu}_{4}}^{t_{2}}$ are mainly composed of $d\gamma$ atom orbitals $\phi_{{\rm Cu}_{(AB)}}^{\gamma}$ at $AB$ sites. The covalent character of the bonding between Cu and Si atoms is weakened, whereas the stability of the Cu$_{4}$ tetrahedron as a molecular unit becomes more important. The details of the molecular-orbital analysis of the Cu$_{4}$ tetrahedron and its relationship with the gap states are given in S4 of Supplemental materials.

It is interesting to compare the electronic structure of $T$-type Cu$_{4}$ with that of other $M_{4}V$  complexes surrounding a vacancy $V$. In this subsection, we deem $T$-type Cu$_{4}$ to be a terminator for a vacancy. This is because this treatment make it easer to compare with the result by Tarnow, who investigated the complexes $M_{4}V$ ($M=$ Li and Br) as the termination of a vacancy.\cite{Tarnow92} 
The details of the electronic structures of $M_{4}V$ ($M$=H, Li, and Br) are given in S5 of Supplemental materials. In all cases, the gap states of a vacancy are completely removed. This is expected from the valence-electron counting method: a vacancy creates six gap states with $t_{2}$ symmetry above the top of the valence band; among them, two states are occupied by electrons, leaving four states empty. Hence, four monovalent atoms can fill the remaining empty states. In this sense, all the $M_{4}V$ complexes examined here have the same role on the terminator of dangling bonds. 

However, the characters of bonding to the host atoms are different and yield secondary effects. 
The best $M_{4}V$ for the purpose of termination is H$_{4}V$. In this case, the four H atoms do not form a dodecahedron: one Si atom has only one H atom as the nearest neighbor with the bond length of 1.5 \AA. A H$_{4}V$ tetrahedron does not behave as a molecular unit because the interaction between H atoms is very weak; thus, the electronic structure can be looked upon as a superposition of four independent Si$-$H bonds.
The next favorable complex is Li$_{4}V$. The Li$-$Si bond slightly elongates the original Si$-$Si bond by 0.05 \AA. In this case, the Li$_{4}$Si$_{4}$ dodecahedron is formed. However, forming a Li$_{4}$ tetrahedron also does not appear to have significant consequence on the valence band of the host crystal except the valence filling, because the electronic charge inside a Li$_{4}$ tetrahedron is almost depleted, as described in Sec.~\ref{sec:chds}. 
Forming a Br$_{4}V$ tetrahedron makes the top of the valence band denser in the DOS, but otherwise no significant change occurs.

For Cu$_{4}V$, the Cu$_{4}$ tetrahedron also acts to terminate a vacancy, but it also has an additional role. The Cu$_{4}$ tetrahedron almost exactly fits the molecular unit within the space of a vacancy, as shown in Sec.~\ref{sec:structure}. The enhancement of the stability of the Cu$_{4}$ tetrahedron by forming a molecular unit may be the cause of the long lifetime of the Cu$_{\rm PL}$ band: the intensity remained above 80\% for 1400 h at room temperature.\cite{Nakamura98}

\section{Phonon properties}
\label{sec:phonons}
The most interesting feature of Cu$_{4}$ is the Cu$_{\rm PL}$ band. By inspection of the phonon sidebands, the frequency of the relevant phonon is found as $\omega_{\rm PL} = 56.8 \ {\rm cm^{-1}}$. We first have to identify the $\omega_{\rm PL}$ mode.
Table \ref{tab:phonons} lists the Cu modes obtained by frozen-phonon calculations. 
The low-symmetry structure ($C$-type), which was previously thought to be the cause of the Cu$_{\rm PL}$ band, has many Cu modes. Accordingly, it is difficult to unambiguously identify $\omega_{\rm PL}$ among these modes. Many modes are seen in the study by Carvalho {\it et al.} too, who gave the phonon spectra for $C$-type.\cite{Carvalho11} 
\begin{table}
\caption{Cu-related phonons of $C$- and $T$-type Cu$_{4}$. The magnitude of the Cu components relative to the normal mode amplitude is given by $c^{2}=\sum_{j={\rm Cu}} u(q|j)^{2}/u(q)^{2}$, where $u(q|j)$ is the $j$th-atom component of $q$th phonon mode $u(q)$. The modes with $c^{2}>10\%$ are retrieved as Cu modes. 
}
\begin{ruledtabular}
\begin{tabular}{c cc| ccc}
  & \multicolumn{2}{c|}{$T$-type} & \multicolumn{3}{c}{$C$-type} \\ \cline{1-6}
   & $\omega$ (${\rm cm^{-1}}$) & $c^{2}$ (\%) & & $\omega$ (${\rm cm^{-1}}$) & $c^{2}$ (\%) \\
 \hline
  $T_{2}$ & 50.4 & 22 & $A_{1}$ & 71.4 & 59   \\
    & 139.4  & 29 &  $E$ & 72.3 & 38  \\
   &  &  & $A_{2}$ & 51.7 & 55 \\
   &  &  & $E$ &  51.0 & 38 \\ 
 \hline
  High freqs. & \multicolumn{2}{l |}{($A_{1}$) 211.5, \quad ($E$) 173.8 }  
  & \multicolumn{3}{c}{Many modes for $\omega > 100 \ {\rm cm^{-1}}$}  \\
   & \multicolumn{2}{l |}{  ($T_{1}$) 155.8}  
  &  & & 
\end{tabular}
\end{ruledtabular}
\label{tab:phonons}
\end{table}%

However, now that the true structure of Cu$_{4}$ has been found to be the high-symmetry form ($T$-type), there is no difficulty in identifying $\omega_{\rm PL}$. As an isolated molecule, the Cu$_{4}$ tetrahedron has the molecular vibrations of five symmetry modes: $A_{1} + E + T_{1} + 2T_{2}$. These include pure translation ($T_{2}$) and pure rotation ($T_{1}$). The corresponding five modes are identified in the present supercell calculation for $T$-type.
There is only one low-frequency mode, which appears at 50.4 ${\rm cm^{-1}}$. This is the translational mode $T_{2,t}$ of the Cu$_{4}$ tetrahedron.
The calculated frequency is close to the observed value of $\omega_{\rm PL} = 56.8 \ {\rm cm^{-1}}$. By observing that there are no other low-frequency modes, the agreement of the frequency with this degree of accuracy is sufficient to identify the $\omega_{\rm PL}$ mode as the $T_{2,t}$ mode of $T$-type.

\begin{figure}[htbp]
      \centering
\includegraphics[width=6.8 cm]{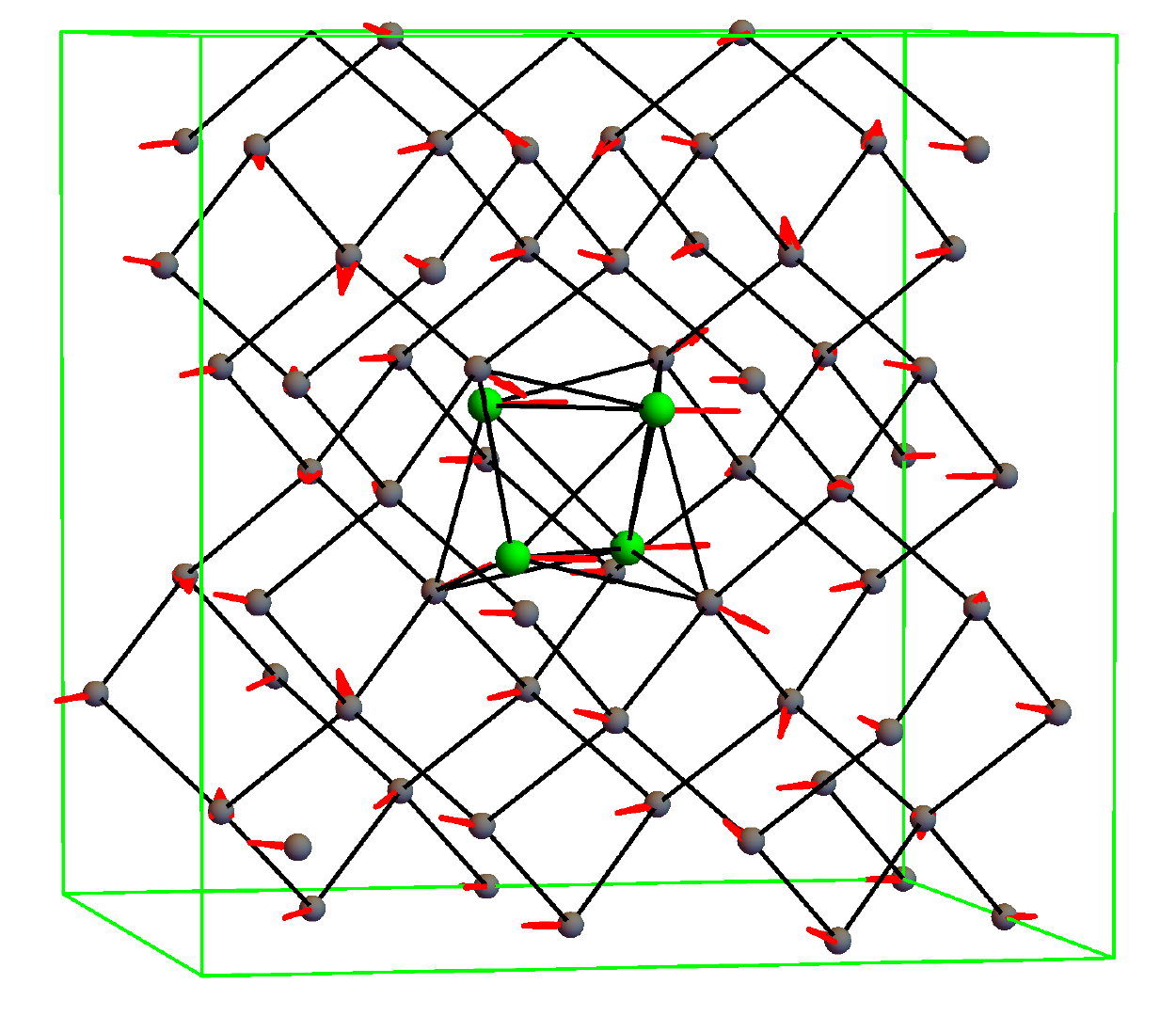}
\hspace{1.0 cm}
\includegraphics[width=7.0 cm]{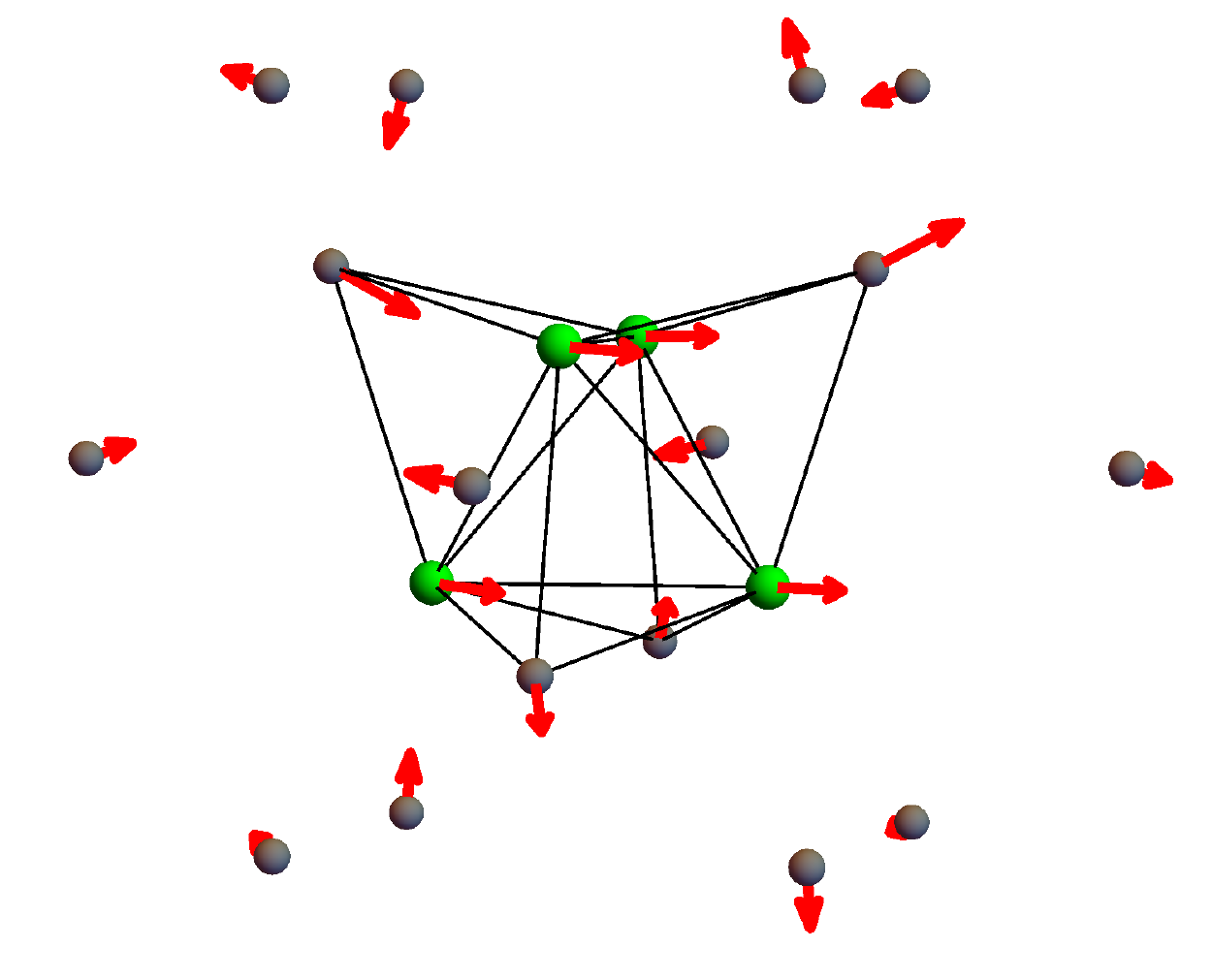} 
\caption{
$T_{2,t}$ mode ($50.4 \ {\rm cm^{-1}}$) of $T$-type Cu$_{4}$. The displacements of all atoms are shown in the left figure. An expanded view around the Cu$_{4}$Si$_{4}$ dodecahedron is shown in the right figure. Green spheres denote Cu and gray ones denote Si.} 
\label{fig:phon-30}
\end{figure}

The $T_{2,t}$ mode for $T$-type Cu$_{4}$ is displayed in Fig.~\ref{fig:phon-30}. 
It is interesting to observe the opposite displacements between Si and Cu atoms.
We call this ``backflow" mode. The backflow mode can be seen as rigid shear translations between the Si and Cu atoms, causing rotational displacements between these two parts in order to make the restoring force minimum.
The contrast of the two flows explains why the rigid translational displacement of Cu$_{4}$ retains its identity in an otherwise homogeneous long-wave mode.



\section{Conclusions}
It has been revealed that the lowest-energy state of Cu$_{4}$ is $T$-type. Its energy is 0.26 eV lower than that of $C$-type, which was previously thought to be the lowest-energy structure. $C$-type, however, remains in a metastable state that is protected from collapsing to $T$-type by an energy barrier of 0.14 eV. 
The mixing of many molecular orbitals of the Cu$_{4}$ tetrahedron with the valence bands eliminates the highly orientated character of covalent bonds, rendering them metallic. The specific geometry of the Cu$_{4}$Si$_{4}$ dodecahedron allows $d\gamma$ orbitals of Cu$_{(i)}$ to couple to the host valence band, giving rise to a gap state just above the top of the valence band. The gap state is closer to the valence top than that of $C$-type. It is likely that Cu$_{\rm PL}$ excitation is caused by excitation of this localized state. The phonon that is responsible to the Cu$_{\rm PL}$ band is likely ascribed to the backflow mode of the Cu$_{4}$ tetrahedron.

The main difficulty in explaining the observed PL spectrum is the symmetry inconsistency: the Cu$_{4}$ tetrahedron has higher symmetry than the observed symmetry of the Cu$_{\rm PL}$ band. The authors hypothesize that this can be explained by Jahn--Teller distortion during the light irradiation in PL measurement. This will be described in a future paper.

\section*{Acknowledgments}
The authors acknowledge M. Nakamura (Ibaraki Univ.) and N. Yarykin (Russian Academy of Sciences) for useful discussion on the experiment on the Cu$_{\rm PL}$ center. They also acknowledge K. Sueoka (Okayama Pref. Univ.) for resolving an issue in the DFT calculation. This study was partly supported by financial support from the Research Program of ``Five-star Alliance” in ``NJRC Mater. \& Dev”.




\end{document}